\documentclass[11pt]{revtex4}
\usepackage{graphics}
\usepackage{amsmath}

\begin{document}
\title{Fitting the WHOIS Internet data}
\author{R. M. D'Souza$^{\dagger}$}
\author{C. Borgs$^{*}$}
\author{J. T. Chayes$^{*}$}
\author{N. Berger$^{\ddagger}$}
\author{R. D. Kleinberg$^{+}$}
\affiliation{$^{\dagger}$\hbox{Dept. of Mechanical and Aeronautical Eng., University of California, Davis}\\ $^{*}$Microsoft Research, Redmond, WA \\
\hbox{$^{\ddagger}$Department of Mathematics, University of California, Los Angeles} \\
\hbox{$^{+}$Department of Computer Science, Cornell University, Ithaca NY}}

\maketitle
This short technical manuscript contains supporting information for Ref.~\cite{dsouza-pnas07}.  We consider the RIPE WHOIS internet data as characterized by the Cooperative Association for Internet Data Analysis (CAIDA)~\cite{CAIDA-3views}, and show that the Tempered Preferential Attachment (TPA) model~\cite{dsouza-pnas07} provides an excellent fit to this data.  First we define the complementary cumulative probability distribution (ccdf), and then derive the ccdf for a TPA graph.  Next we discuss the ccdf for the WHOIS data. Finally we discuss the fit provided by the TPA model and by a power law with exponential decay (PLED). 

\section{Defining the CCDF}
The complementary cumulative probability distribution, ccdf$(x)$:
\begin{equation}
{\rm ccdf}(x) = 1 - \sum_{j=1}^{x-1} p_j = \sum_{j=x}^{\infty} p_j.
\label{eqn:ccdf}
\end{equation}

\section{The CCDF predicted by TPA with $A_1 \ne A_2$}
\subsection{First recall the recursion relations} 
The recursion relations defining the degree distribution for TPA graphs were derived explicitly in Refs.~\cite{CIPA04} and \cite{CIPA05}.  Here we derive the corresponding ccdf.   These are Eqn's (16) and (17) in~\cite{CIPA04}:
\begin{equation}
p_i = \left( \prod_{k=2}^i \frac{k-1}{k+w}\right) p_1 = \left( \prod_{k=1}^{i-1} \frac{k}{k+w+1}\right) p_1, \ \ \ {\rm for} \ \ i \le A_2,
\end{equation}
and
\begin{equation}
p_i = \left( \frac{A_2}{A_2+w}\right)^{i-A_2} p_{A_2} \ = \  q^{i-A_2}\  p_{A_2}, \ \ \ {\rm for} \ \ i \ge A_2.
\label{eqn:plessA2}
\end{equation}
Note
\begin{equation}
p_{A_2} = \left( \prod_{k=1}^{A_2-1} \frac{k}{k+w+1}\right) p_1,
\end{equation}
and, for convenience, we defined: 
\begin{equation}
q \equiv \left( \frac{A_2}{A_2+w} \right).
\end{equation}

We will first calculate the CCDF for $i \ge A_2$ as we will use that result to determine the CCDF for $i < A_2$. 

\subsection{Calculating the CCDF, for $x \ge A_2$}
Recall the definition of the CCDF from Eqn.~(\ref{eqn:ccdf}):
\begin{eqnarray}
{\rm ccdf}(x) & = & \sum_{j=x}^\infty p_j \nonumber \\
& = & p_{A_2} \sum_{j=x}^\infty q^{j- A_2} \nonumber \\
& = & p_{A_2} \sum_{j=0}^\infty q^{j + x - A_2} \nonumber \\
& = & p_{A_2} q^{x-A_2} \sum_{j=0}^\infty q^j. \label{eqn:geom}
\end{eqnarray}

Since $q<1$, the sum in Eqn.~(\ref{eqn:geom}) is a geometric series; $\sum_{j=0}^\infty q^j = 1/(1-q)$. Thus we can write:

\begin{equation}
\boxed{{\rm ccdf}(x) = \left(\frac{p_{A_2}}{1-q}\right) q^{x-A_2}, \ \ {\rm for} \ \ x \ge A_2.}
\label{eqn:ccdfgreaterA2}
\end{equation}

\subsection{Calculating the CCDF, for $x < A_2$}
This is slightly more complicated, as we have different functional forms for $x< A_2$ and $x > A_2$.

\begin{eqnarray}
{\rm ccdf}(x) & = & \sum_{j=x}^{\infty} p_j \nonumber \\
 & = & \sum_{j=x}^{A_2-1} p_j + \sum_{j = A_2}^\infty p_j \nonumber \\
 & = & \sum_{j=x}^{A_2-1} p_j + {\rm ccdf}(A_2) \nonumber \\
 & = & \sum_{j=x}^{A_2-1} p_j + \left(\frac{p_{A_2}}{1-q}\right).
\end{eqnarray}

Plugging in the relation for $p_i$ from Eqn.~(\ref{eqn:plessA2}), we obtain:
\begin{equation}
\boxed{
{\rm ccdf}(x) = p_{A_2} \left(\frac{1}{1-q} + \sum_{j=x}^{A_2-1}  \prod_{k=j}^{A_2 -1} \frac{k+w+1}{k}\right),  \ \ {\rm for} \ \ x < A_2.
}
\label{eqn:ccdflessA2}
\end{equation}

\subsection{Standard Normalization}
First we can check that Eqns.~(\ref{eqn:ccdfgreaterA2}) and~(\ref{eqn:ccdflessA2}) give the same value for ccdf$(A_2)$. They do:
\begin{equation}
{\rm ccdf}(A_2) = \frac{p_{A_2}}{1-q}.
\end{equation}
And we can determine the value of $p_{A_2}$ by the normalization condition that
\begin{equation}
{\rm ccdf}(1) = 1 =  p_{A_2} \left(\frac{1}{1-q} + \sum_{j=1}^{A_2-1}  \prod_{k=j}^{A_2 -1} \frac{k+w+1}{k}\right).
\end{equation}
In other words, 
\begin{equation}
\boxed{
p_{A_2} = \left(\frac{1}{1-q} + \sum_{j=1}^{A_2-1}  \prod_{k=j}^{A_2 -1} \frac{k+w+1}{k}\right)^{-1}.
}
\end{equation}

\subsection{Normalizing without degree $d=1$ nodes}
We may want to neglect nodes with degree $d<2$ for various reasons.  In that case, the normalization would be:
\begin{equation}
{\rm ccdf}(2) = 1 =  p_{A_2} \left(\frac{1}{1-q} + \sum_{j=2}^{A_2-1}  \prod_{k=j}^{A_2 -1} \frac{k+w+1}{k}\right).
\end{equation}
Thus
\begin{equation}
\boxed{
p_{A_2} = \left(\frac{1}{1-q} + \sum_{j=2}^{A_2-1}  \prod_{k=j}^{A_2 -1} \frac{k+w+1}{k}\right)^{-1}
}
\label{eqn:pA2}
\end{equation}
with Eqns.~(\ref{eqn:ccdfgreaterA2}) and~(\ref{eqn:ccdflessA2}) unchanged (except Eqn.~(\ref{eqn:ccdflessA2}) now holds for $2 \le x < A_2$, rather than for $1 \le x < A_2$).

\section{The WHOIS ccdf, for $d>1$}
\subsection{Whois data, renormalize to remove $d<2$}
By definition: $$\sum_{j=1}^{\infty} p_j =1.$$

Thus: $$\sum_{j=2}^{\infty} p_j = 1 - p_1.$$

We want to renormalize ($p_j' = \eta p_j$) such that: $$ \sum_{j=2}^{\infty} p_j' =  \eta \sum_{j=2}^{\infty} p_j = 1,$$

Thus \fbox{$\eta = 1/(1-p_1).$} For the Whois data, $p_1 = 0.0573.$ and $\eta = 1.0608.$

The {\bf complementary cumulative distribution function} (ccdf) for the renormalized probabilities:
$$\rm{ccdf'}(x) = \sum_{j=x}^{\infty} p_j' = \eta \sum_{j=x}^{\infty} p_j = \eta \  \rm{ccdf}(x).$$
\begin{figure}[b]
\resizebox{3in}{!}{\includegraphics{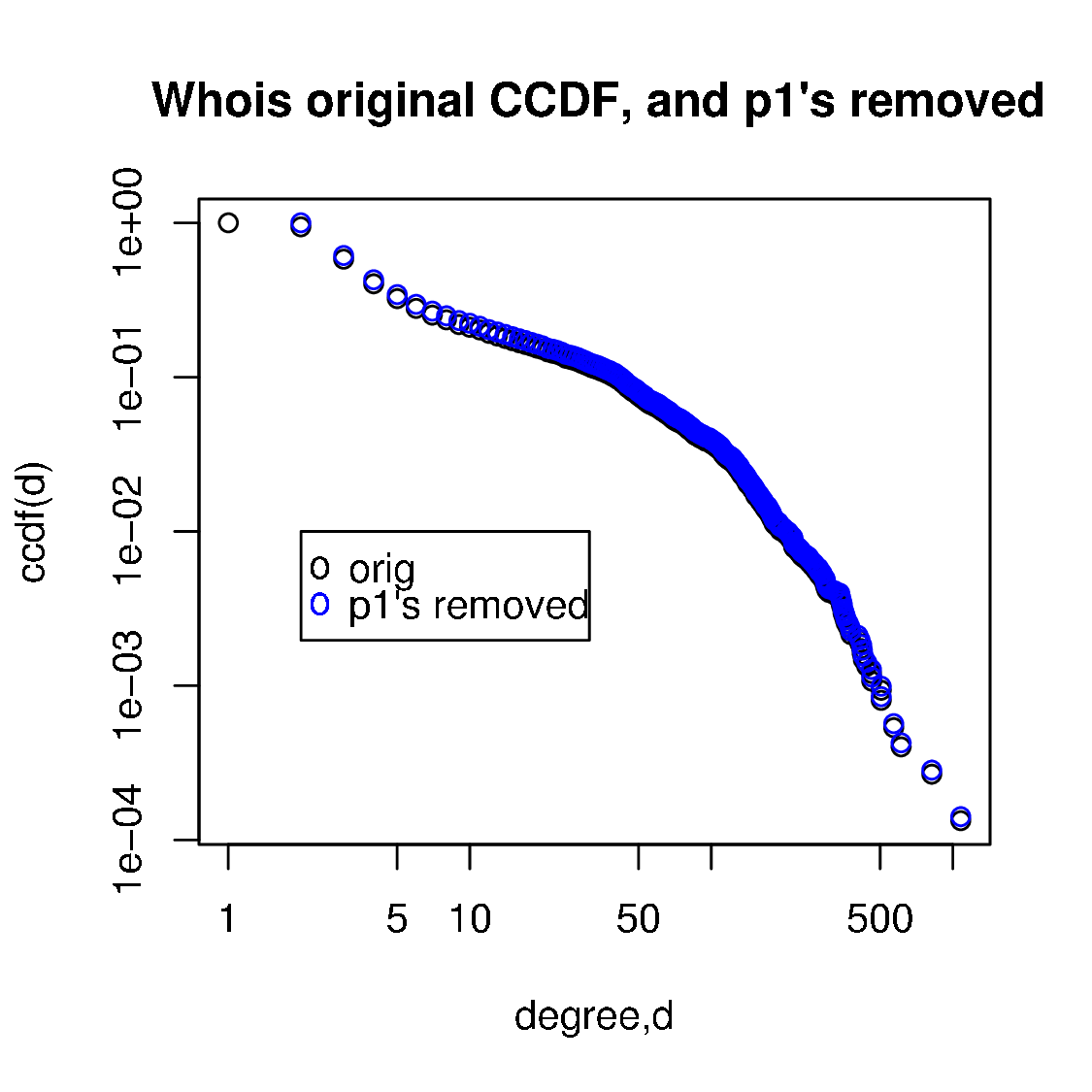}}
\caption{Original CCDF of Whois data, and the renormalized CCDF'(x) $= \eta$ CCDF(x).}
\end{figure}

\section{Fitting TPA to WHOIS with $d\ge 2$}
Whois $d \ge 2$ distribution discussed above.  TPA with $d\ge 2$ is the same as with $d \ge 1$ except the value of $p_{A_2}$ is defined as in Eqn.~(\ref{eqn:pA2}), in terms of $d=2$ instead of $d=1$. 
\begin{figure}[h]
\resizebox{3in}{!}{\includegraphics{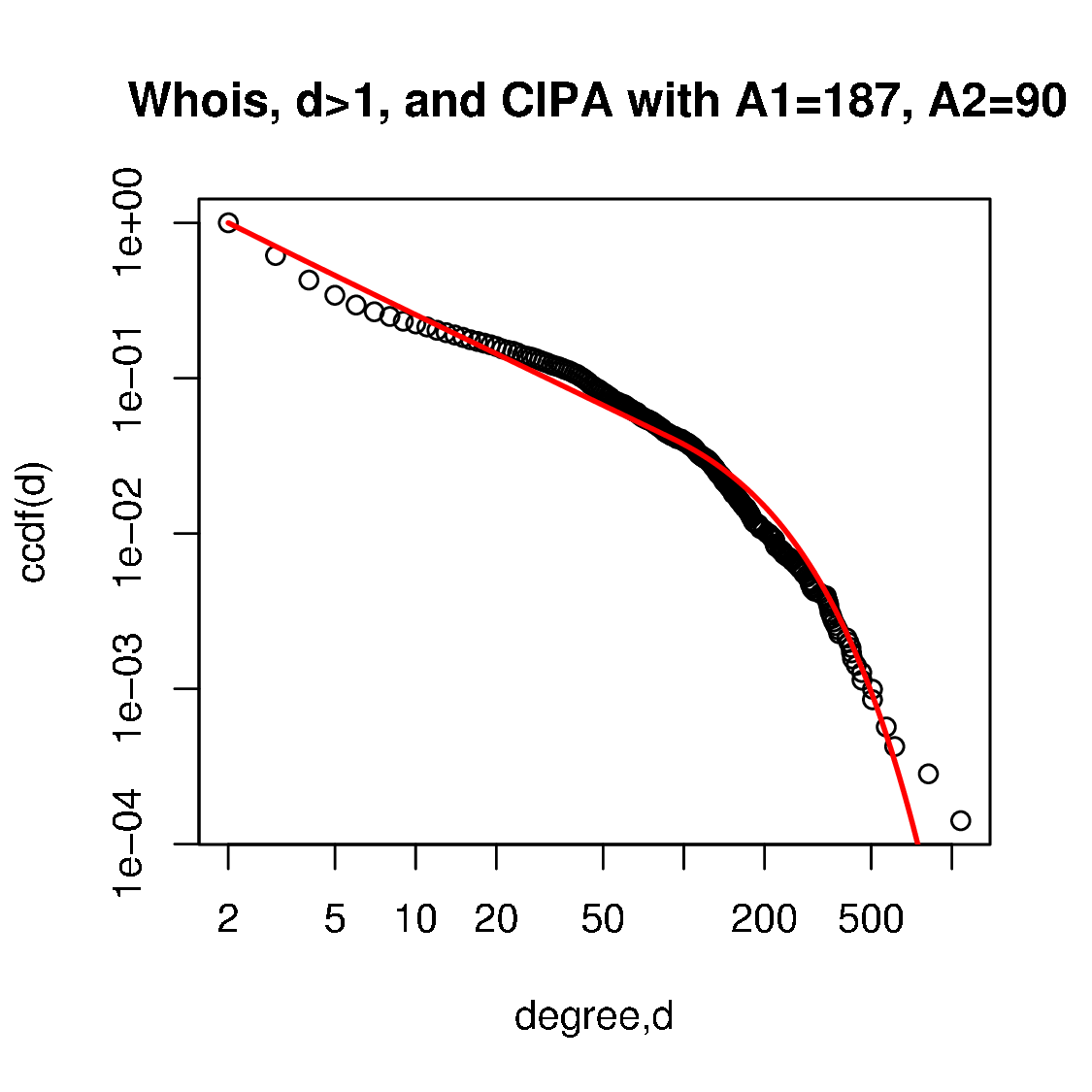}}
\vspace{-0.1in}
\caption{Whois CCDF for $d \ge 2$.  Data points are from the Whois tables. The solid line is the fit to TPA for $d \ge 2$ with $A_1=187$ and  $A_2=90$ (and thus $\gamma = 1.83$). With this fit, $R=0.986$, thus $R^2 = 0.972$.}
\end{figure}


\section{Fitting PLED to WHOIS with $d\ge 2$}

Assuming a PLED: $p(x) = A x^{-b} \exp(-x/c)$.  The normalization constant, $A$, is determined by the relation:
$$ \sum_{x=2}^{\infty} p(x) = 1 = A \sum_{x=2}^{\infty} x^{-b} \exp(-x/c).$$

Then the ccdf:
$${\rm ccdf}(x) = A \sum_{j=x}^\infty x^{-b} \exp(-x/c).$$

\begin{figure}[h]
\resizebox{3in}{!}{\includegraphics{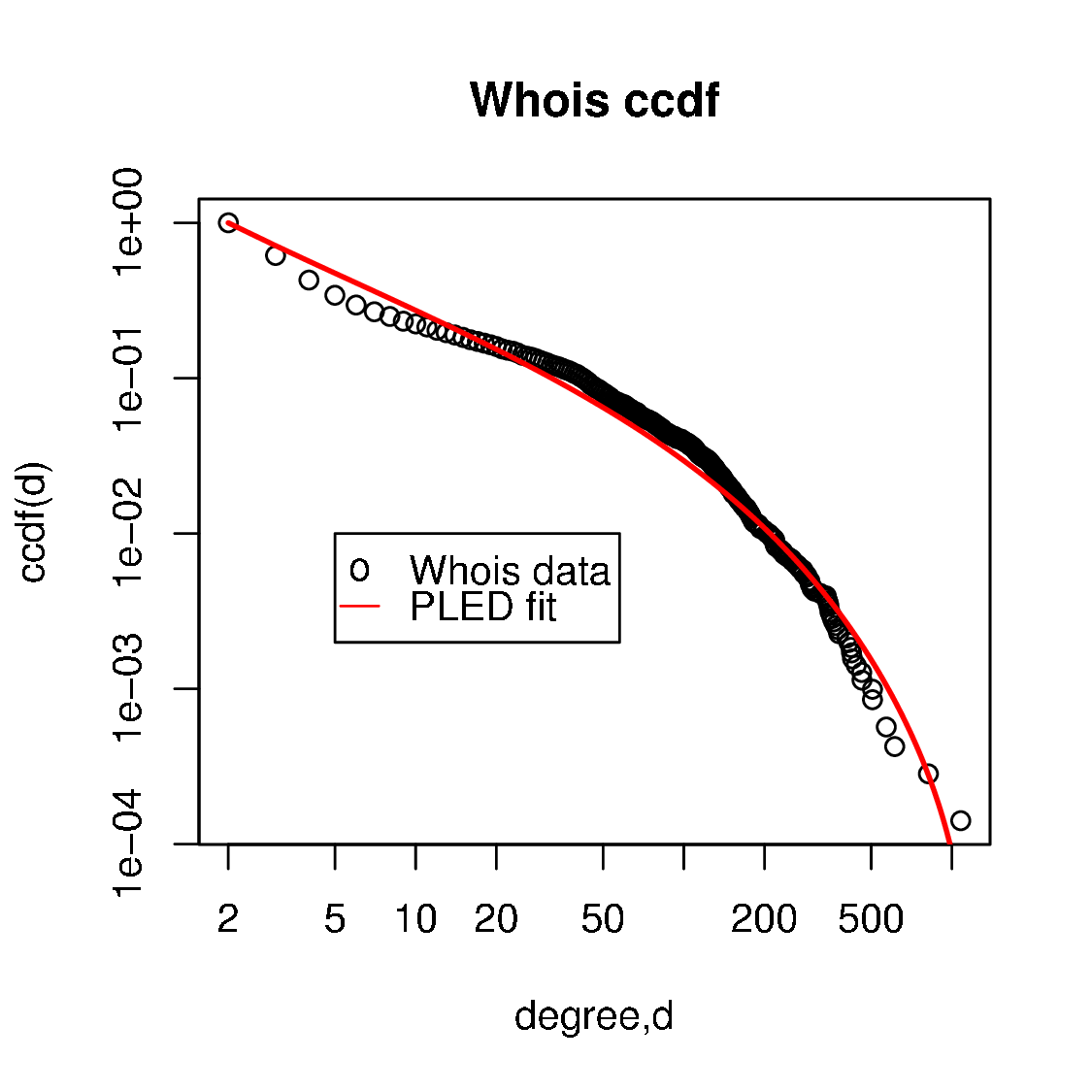}}
\vspace{-0.1in}
\caption{Whois CCDF for $d \ge 2$.  Data points are from the Whois tables. The solid line is the fit ${\rm ccdf}(x) = A \sum_{j=x}^\infty x^{-b} \exp(-x/c)$, where $b=1.63$ and $c=350$.  With this fit, $R=0.985$, thus $R^2 = 0.970$.}
\end{figure}

\end{document}